\documentstyle[eqsecnum,preprint,aps,epsf]{revtex}
\tightenlines
\input epsf
\newcommand{\be}{\begin{equation}}
\newcommand{\ee}{\end{equation}}
\newcommand{\bea}{\begin{eqnarray}}
\newcommand{\eea}{\end{eqnarray}}
\newcommand{\OP}{\vec{\psi}}
\newcommand{\SG}{\vec{\sigma}}
\newcommand{\M}{\vec{m}}

\newcommand{\F}{{\cal F}}

\newcommand{\Q}{Q_{\alpha \beta}}
\newcommand{\N}{\hat{n}}
\newcommand{\HM}{\hat{m}}
\newcommand{\SV}{\vec{s}}
\begin{document}
\draft
\title{Phase ordering in bulk uniaxial nematic liquid crystals}
\author{Robert A. Wickham}
\address{The James Franck Institute and Department of Physics\\ The
University of Chicago\\ Chicago, Illinois 60637}
\date{\today}
\maketitle
%
% ABSTRACT
%
\begin{abstract}
The phase-ordering kinetics of a bulk uniaxial nematic liquid crystal
is addressed using techniques that have been successfully applied to 
describe ordering in the $O(n)$ model. The method involves constructing 
an appropriate mapping between the order-parameter tensor and a Gaussian
auxiliary 
field. The mapping accounts both for the geometry of the director about the 
dominant charge 1/2 string defects and biaxiality near the string cores. 
At late-times $t$ following a quench, there exists a scaling regime where 
the bulk nematic liquid crystal and the three-dimensional $O(2)$ model 
are found to be isomorphic, within the Gaussian approximation. 
As a consequence, the scaling function for 
order-parameter correlations in the nematic liquid crystal 
is {\em exactly} that of the $O(2)$ model, and the length characteristic of 
the strings grows as $t^{1/2}$. These results are in accord with experiment 
and simulation. Related models dealing with thin films and monopole 
defects in the bulk are presented and discussed.   
\end{abstract}
%
% INTRODUCTION
% 
\section{Introduction}

Most phase-ordering systems studied to date support only one type of 
topologically stable defect species \cite{MAZENKO90,LIU92a,BRAY94}.
One example is the $O(n)$ model with an $n$-component vector order-parameter. 
In three spatial dimensions, the defects formed at the quench 
are line-like strings for $n=2$, and point-like monopoles for $n=3$.
Phase-ordering in bulk uniaxial nematic liquid crystals (nematics) provides
the simplest scenario in which two defect species - monopoles 
and strings - are topologically stable. The stability of monopoles 
derives from the $O(3)$ symmetry of the nematic director $\hat{n}(\vec{r},t)$.
The additional invariance under the local inversion $\hat{n}(\vec{r},t) 
\rightarrow -\hat{n}(\vec{r},t)$ allows the nematic to support 
stable charge $1/2$ disclination lines (strings) \cite{MERMIN79}. 
The issue of which defect species dominates the dynamics in bulk nematics 
at late times $t$ following a quench has recently been settled. 
Cell-dynamical simulations using spin models of bulk nematics 
\cite{BLUNDELL92,TOYOKI94} have computed the order-parameter correlation 
function and found it to be indistinguishable from that of the $O(2)$ model,
and consistent with a string-dominated late-time scaling regime. Experiments 
by Chuang {\em et al.} \cite{CHUANG93} directly imaged the bulk nematic, 
revealing an intricate, evolving defect tangle. 
While both types of defect were observed, the strings dominated at
late times. The length scale $L_{s}$ characterizing the typical 
separation of the strings was seen to
grow as $L_{s} \sim t^{1/2}$ while the average line density of
string $\langle \eta \rangle$
decayed like $\langle \eta \rangle
\sim L_{s}^{-2} \sim t^{-1}$. The study of ordering in 
nematics is also of interest to cosmologists 
\cite{CHUANG91,COSMOS} since similar processes involving cosmic string and 
monopole evolution, thought to occur in the early universe, may be
responsible for structure formation.
 
In this paper a theory is presented that describes the dominant scaling 
behaviour of the bulk nematic in terms of a string-dominated late-time 
regime. Generalizing a successful method used to treat
the ordering kinetics of the $O(n)$ model, the nematic order-parameter 
tensor is mapped  onto a two-component Gaussian auxiliary field \cite{LIU92a}.
The string defects explicitly appear in the construction of the mapping.
As discussed below, this approach has several advantages over an earlier, 
semi-numerical theory by Bray {\em et al.} \cite{BRAY93}.

\begin{figure}
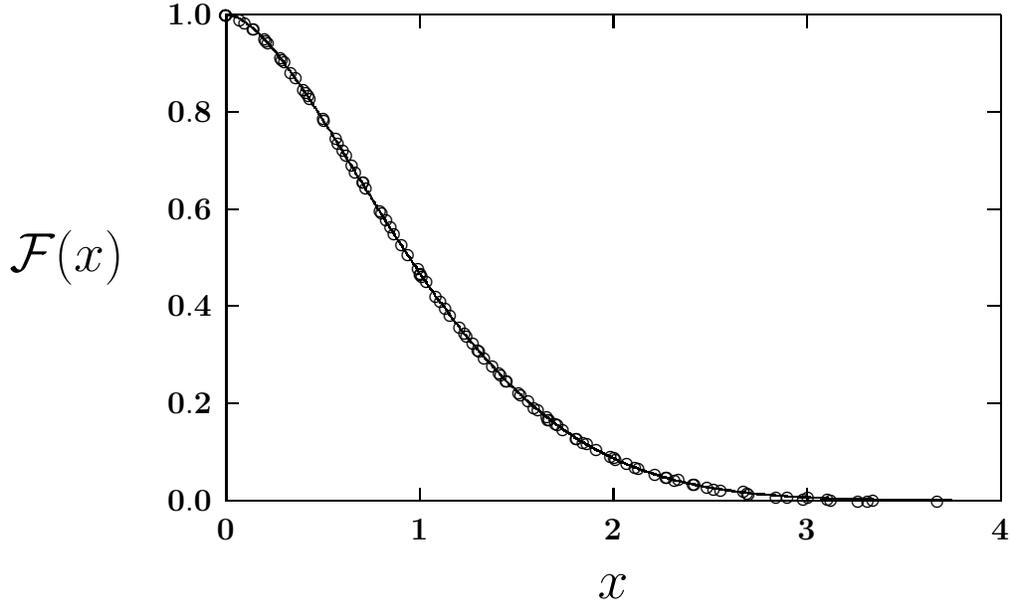

% GNUPLOT: LaTeX picture
\setlength{\unitlength}{0.240900pt}
\ifx\plotpoint\undefined\newsavebox{\plotpoint}\fi
\sbox{\plotpoint}{\rule[-0.200pt]{0.400pt}{0.400pt}}%

\vspace{.5 in}
\caption{The scaling function $\F(x)$ for order-parameter correlations 
as a function of the scaled length $x=r/L_{s}(t)$ for the three dimensional
$O(2)$ model \protect\cite{LIU92a} is represented by the solid line. 
As shown in Section 
III.B this function exactly describes order-parameter correlations in 
bulk nematics. The cell-dynamical simulation data of Blundell and Bray 
\protect\cite{BLUNDELL92} for this quantity in a bulk nematic are also 
shown, as circles. The abscissa of the simulation data is scaled so as to 
give the best fit to the theory.}
\label{FIG:STR}
\end{figure}

The auxiliary field approach is first applied to the straightforward
case of phase-ordering in nematic films containing charge $1/2$
vortices, which have been studied in simulations \cite{BLUNDELL92} 
and experiments \cite{WONG92}. As in the bulk nematic, the 
mapping is constructed to account for the rotation of the director by 
only $\pi$ about the core of the defect. Once this is done the theory 
reveals that phase-ordering in the nematic film is equivalent to phase 
ordering in the two-dimensional $O(2)$ model examined previously 
\cite{LIU92a}. This is not surprising since the two systems are known to 
be isomorphic \cite{BLUNDELL92,DESAI}. Constructing a theory for the 
bulk nematic is more challenging since the order-parameter 
tensor must include a biaxial piece near the core of the string. 
In the earlier theory of Bray {\em et al.} 
\cite{BRAY93} this point was not addressed since there they used 
a ``hard-spin'' approximation for the dynamics of the nematic.
However, the necessity of a having a biaxial core region when treating 
the full equations has been noted in the numerical work of Schopohl and 
Sluckin \cite{SCHOPOHL87} on bulk nematic string defects in equilibrium. 
The present theory successfully incorporates biaxiality and clarifies 
the role that it plays in the coarsening of the bulk nematic. 
The theory recovers the growing length $L_{s} \sim t^{1/2}$ seen in 
simulations \cite{BLUNDELL92,TOYOKI94} and  
experiments \cite{CHUANG93}. 
In the scaling regime, the order-parameter 
correlation function for the bulk nematic is found to be {\em exactly} 
that of the three-dimensional $O(2)$ model \cite{LIU92a}, 
in excellent agreement with simulations \cite{BLUNDELL92} 
(Fig. \ref{FIG:STR}). Although the theoretical results of
Bray {\em et al.} \cite{BRAY93} suggested agreement between the 
correlation function for the bulk nematic and the $O(2)$ model 
they were unable to demonstrate an exact 
equivalence since their theory was not based on a mapping that explicitly 
contained strings. The major accomplishment of this work is to 
analytically demonstrate the 
isomorphism between the dynamics of the bulk nematic and the dynamics of the 
three-dimensional $O(2)$ model, within the Gaussian approximation. 
Through this isomorphism, the well-developed 
theory for the $O(2)$ model \cite{LIU92a,LIU92b,MAZENKO97FL} can be 
applied directly to the nematic. In particular, this theory predicts that
the average line density of string
decays as $\langle \eta \rangle \sim L_{s}^{-2} \sim t^{-1}$ 
\cite{LIU92b,WICKHAM97b}, in accord with the 
experiments of Chuang {\em et al.} \cite{CHUANG93}.

Although strings are generically present in bulk nematics, certain choices of 
experimental setup and sample material will produce copious amounts of 
monopoles at the quench \cite{PARGELLIS96}. The theory of 
Bray {\em et al.} \cite{BRAY93} is unable to address these experiments 
since in that theory there is no signature for monopoles. However,
within the framework presented below it is relatively straightforward to 
develop a  theory of nematics in which monopoles appear. 
In this theory the order-parameter correlation function 
is found to be similar to 
(but not exactly) that for the three-dimensional $O(3)$ model \cite{LIU92a}
(Fig. \ref{FIG:MON}). 
The characteristic monopole spacing $L_{m}$ grows as $L_{m} \sim t^{1/2}$ 
and leads to a decaying average monopole density $\langle n \rangle
\sim L_{m}^{-3} \sim t^{-3/2}$. Experiments \cite{PARGELLIS96} that 
examine monopole-antimonopole annihilation in isolation from strings 
suggest that these growth laws should hold. However, experiments
\cite{CHUANG93} also reveal that the average
monopole density decays more rapidly
in the presence of strings, with $\langle n \rangle 
\sim t^{-3}$. It appears that in order
to account for this observation the theory presented here should be extended 
to consider the interactions between strings and monopoles \cite{WICKHAM97b}.  
%
% MODELS
%
\section{Models}

In this section the $O(n)$ model and the Landau-de Gennes model of nematics 
are discussed. Since the former model is used as a guide in the
treatment of the latter, the theory for  
ordering kinetics in the $O(n)$ model is also reviewed. 
Initially, the structural features 
common to both models are emphasized. In later sections, 
the technical details specific to the ordering of nematics will be discussed. 
%
% THE O(N) MODEL
%
\subsection{The $O(n)$ model}

In the $O(n)$ model the evolution of the non-conserved, $n$-component 
order-parameter field $\OP$ is governed by the time-dependent 
Ginzburg-Landau (TDGL) equation
%
% LANGEVIN EQUATION
\be
\frac{\partial \OP}{\partial t} = - \frac{\delta F[\OP]}{\delta \OP}.
\label{EQ:TDGL}
\ee
The free energy $F[\OP]$ has the form
%
% FREE ENERGY
\be
\label{EQ:ONFREE}
F[\OP] = \int d^{d} r \mbox{ } [ \frac{1}{2} (\nabla \OP)^{2} + V(\psi)]
\ee
where the potential $V(\psi)$, expressed in terms of $\psi \equiv  |\OP|$,
is $O(n)$ symmetric with a degenerate ground state at non-zero 
$\psi = \psi_{0}$. In this model, as with the nematic liquid crystal, 
the disordered high temperature initial state is rendered unstable by a 
quench to a low temperature where the usual noise term on the right-hand 
side of (\ref{EQ:TDGL}) can be ignored. Substitution of  (\ref{EQ:ONFREE}) 
into (\ref{EQ:TDGL}) produces the explicit equation of motion  
%
% EQUATION OF MOTION
\be
\frac{\partial \OP}{\partial t} = \nabla^{2} \OP - 
\frac{\partial V(\psi)}{\partial \OP}.
\label{EQ:MOT}
\ee

The evolution induced by (\ref{EQ:MOT}) causes $\OP$ to order and assume 
a distribution that is far from Gaussian. To make analytic progress it
is by now standard \cite{MAZENKO90} to introduce a mapping
%
% MAPPING
\be
\OP = \SG ( \M )
\ee
between the physical field $\OP$ and an $n$-component 
auxiliary field $\M$ with analytically tractable statistics.
The mapping $\SG$ is chosen to reflect the defect structure in the system
and satisfies the Euler-Lagrange equation for 
a defect in equilibrium 
%
% RELATION TO POTENTIAL
\be
\nabla^{2}_{m} \SG = \frac{\partial V (\SG)}{\partial \SG}.
\label{EQ:SIGPOT}
\ee
As shown below, (\ref{EQ:SIGPOT}) is also instrumental in 
treating the non-linear potential term in (\ref{EQ:MOT}).
Defects correspond to the non-uniform solutions of (\ref{EQ:SIGPOT}) which 
match on to the uniform solution far from the defect core. Since only the 
lowest-energy defects, those with unit topological charge, will survive 
until late-times the relevant 
solutions to (\ref{EQ:SIGPOT}) will be of the form \cite{LIU92a}
%
% EXPLICIT FORM FOR SIGMA
\be 
\SG (\M) = A(m) \hat{m}
\label{EQ:SIGEX}
\ee
where $m = |\M| \mbox{ and } \hat{m} = \M/m$. 
Thus the magnitude of $\M$ represents the distance away from a defect core and
its orientation corresponds to the orientation of the order parameter field 
at that point. This geometrical interpretation will later be exploited 
when the generalization of (\ref{EQ:SIGPOT}) is used to choose the 
appropriate mapping, analogous to (\ref{EQ:SIGEX}), for string defects in the 
nematic liquid crystal. The magnitude of $\M$ grows as the characteristic 
defect separation, $L(t)$, becoming large in the late-time, scaling regime.
Inserting (\ref{EQ:SIGEX}) into (\ref{EQ:SIGPOT}) gives an equation for $A$,
the order-parameter profile around a defect \cite{LIU92a}
%
% `A' EQUATION 
\be
\nabla^{2}_{m} A - \frac{n-1}{m^{2}} A - \frac{\partial V}{\partial A}(A) = 0.
\label{EQ:A}
\ee
For small $m$ an analysis of 
(\ref{EQ:A}) yields the linear dependence $A(m) \sim m$, characteristic of 
a unit charge defect \cite{CHARGE}. 
For large $m$ the amplitude $A$ approaches its ordered value 
$A=\psi_{0}$ algebraically, which is a feature 
common to both the $O(n)$ model and the nematic.
                                  
The order-parameter correlation function is
%
% ORDER PARAMETER CORRELATION FUNCTION
\be
\label{EQ:OOOP}
C(\vec{r},t) = \langle \SG (\vec{r},t) \cdot \SG (0,t) \rangle = 
\psi_{0}^2 \langle \hat{m}(\vec{r},t) \cdot \hat{m}(0,t) \rangle
\ee
where the last equality holds for late-times and to leading order in $1/L$.
To evaluate the last average in (\ref{EQ:OOOP}) we choose $\M$ to be a
Gaussian field with zero mean. 
This Gaussian approximation forms the basis of almost 
all present analytical treatments of phase-ordering problems, 
and has had much quantitative success in describing the correlations in these 
systems \cite{MAZENKO90,LIU92a,BRAY94}. Theories where $\M$ is a 
non-Gaussian field also exist \cite{MAZENKO94a,WICKHAM97a}. 
In the Gaussian approximation the order-parameter correlation function 
(\ref{EQ:OOOP}) can be related to the normalized auxiliary field
correlation function $f$, defined as 
%
% DEFINITION OF f
\be
f(\vec{r},t) \equiv \frac{\langle \M (\vec{r},t) \cdot \M (0,t) \rangle}
                    {\langle [\M(0,t)]^{2} \rangle}.
\label{EQ:DEFNF}
\ee
The relation is \cite{LIU92a,BRAY91}
%
% GAUSSIAN FORM FOR C
\be
\label{EQ:GAUSSIANC}
C(\vec{r},t) = \psi_{0}^{2} \F (\vec{r},t)
\ee
with 
\be
\label{EQ:ONSCALF}
\F = \frac{nf}{2 \pi} B^{2} \left[ \frac{1}{2},\frac{n+1}{2} \right]
F \left[ \frac{1}{2}, \frac{1}{2};\frac{n+2}{2};f^{2} \right],
\ee
where $B$ is the beta function and $F$ is the hypergeometric function.
In the late-time scaling regime the functions 
$\F$ and $f$ can be expressed solely in terms of the scaled length 
$x=r/L(t)$ so that $\F=\F(x)$. 
In this regime the equation of motion (\ref{EQ:MOT})
can be written as non-linear scaling equation for 
$\F$ 
%
% GAUSSIAN EQUATION FOR F
\be
\vec{x} \cdot \nabla \F + \nabla_{x}^{2} \F 
+ \frac{\pi}{2 \mu} f \frac{\partial}{\partial f} \F = 0.
\label{EQ:G-EQNF}
\ee
In the derivation of (\ref{EQ:G-EQNF}) the relation (\ref{EQ:SIGPOT}) is 
used to replace the potential term in (\ref{EQ:MOT}), 
and then the Gaussian identity
%
% GAUSSIAN IDENTITY
%
\be
\label{EQ:GI}
\langle [\nabla_{m}^{2} \SG (\M(\vec{r},t))] \cdot \SG (\M(0,t)) \rangle =
- \frac{n f(\vec{r},t)}{\langle  [\M(0,t)]^{2} \rangle} 
\frac{\partial}{\partial f(\vec{r},t)} \langle \SG(\M(\vec{r},t)) \cdot
\SG(\M(0,t)) \rangle
\ee
is used to get the last term on the left-hand side of  (\ref{EQ:G-EQNF}).
The constant $\mu$ enters through the definition of the scaling length $L$:
%
% DEFINITION OF MU
\be
L^{2} (t) \equiv \frac{\pi \langle [\M(0,t)]^{2} \rangle }{2 n \mu} = 4 t.
\label{EQ:DEFNMU}
\ee
This is  the well-known \cite{LIU92a,BRAY91} growth law $L \sim t^{1/2}$ 
for phase-ordering in non-conserved vector systems.

Since the auxiliary field $\M$ is smooth \cite{MAZENKO97FL}, 
$f$ is analytic for small-$x$. 
This implies, through an  examination of  (\ref{EQ:G-EQNF}) in $d$ spatial
dimensions, that for small-$x$ $\F$ behaves like 
%
% GAUSSIAN BOUNDARY CONDITIONS - SHORT DISTANCE
\be
\F(x) = 1 + \frac{\pi}{4 \mu d} x^{2} \ln x + {\cal O} (x^2)
\label{EQ:GS}
\ee
for $n = 2$ and 
\be
\F(x) = 1 - \frac{\pi}{2 \mu d}  x^{2} + \frac{4}{3 \mu (d+1)} \sqrt{\frac{\pi}
{2 \mu d}} \mbox{ } x^{3} +  {\cal O} (x^{4})
\label{EQ:GS1}
\ee
for $n=3$, the cases relevant to this paper. The non-analytic terms in $\F$ 
reflect the short-distance singularities in the order 
parameter field produced by the defects, and lead to the Porod's law 
\cite{POROD} power law decay of the structure factor at large wavenumber. 
The  $x^{2} \ln x$ term in (\ref{EQ:GS}) is characteristic of string (or 
vortex) defects while the $x^{3}$ term in (\ref{EQ:GS1})
is due to monopole defects. 
For large $x$ both $\F$ and $f$ decay rapidly to zero.
 The eigenvalue $\mu$ is determined numerically by 
matching the short- and long-distance behaviours of the solution of
(\ref{EQ:G-EQNF}). In this way the auxiliary field correlation function $f$ is
determined self-consistently along with $\F$. In contrast, there is no
such self-consistency in theories based on the 
Ohta-Jasnow-Kawasaki (OJK) approximation \cite{BRAY93,OJK}. 
Values of $\mu$ at various $n$ and $d$ for the 
$O(n)$ model have been determined \cite{LIU92a}. 
The scaling functions $\F$ of this theory are in excellent agreement 
with the results of simulations \cite{MAZENKO90,LIU92a}.
%
% NEMATIC LIQUID CRYSTALS
%
\subsection{Nematic Liquid Crystals}

The order-parameter for a bulk nematic liquid crystal is a  traceless, 
symmetric, $3 \times 3$ tensor $\Q$, which measures the anisotropy of 
physical observables in the nematic phase. 
The tensor $\Q$ has the general form 
\cite{PONIEWIERSKI85}
%
% Q TENSOR - 3D
%
\be
\label{EQ:3DQ}
\Q = A [ \hat{n}_{\alpha} \hat{n}_{\beta} - \frac{1}{3} \delta_{\alpha \beta} ]
+ \frac{1}{3} B [ \hat{g}_{\alpha} \hat{g}_{\beta} - \hat{h}_{\alpha} 
\hat{h}_{\beta}].
\ee
The unit $3$-vectors $\hat{n}$, $\hat{g}$ and $\hat{h}$ form an 
orthonormal triad. The amplitudes 
$A$ and $B$ are chosen to be non-negative.
$A$ is a measure of the degree of uniaxial order in 
the liquid crystal; it is zero in the isotropic phase and non-zero 
in the nematic phase. Biaxiality in the liquid crystal is measured by $B$.
In the uniaxial 
nematic
phase $B$ is zero everywhere except near the string cores. The description of 
nematics in terms of $\Q$ reduces to the Frank continuum theory of elasticity 
in terms of a director $\hat{n}(\vec{r},t)$ 
\cite{DEGENNES} when $A$ is set to its ordered value and $B=0$. 
In the phase-ordering scenario, where defects occur, all of $A$, $B$, 
$\hat{n}$, $\hat{g}$ and $\hat{h}$ are space and time dependent.

In the tensor formulation the director, 
which measures the average local molecular orientation in the nematic,
is the unit eigenvector of $\Q$ which corresponds to the largest eigenvalue.
The unit eigenvectors and associated eigenvalues of 
$\Q$ are:
%
% ARRAY OF E_VALS AND VECTORS
%
\be
\begin{array}{ccc}
\hat{n} & \leftrightarrow &  \frac{2}{3} A \\
\hat{g} & \leftrightarrow & -\frac{1}{3} (A-B) \\
\hat{h} & \leftrightarrow & -\frac{1}{3} (A+B) 
\end{array}
\ee
Since the nematic is uniaxial, $B \leq 3A$ and 
the director can be identified with $\hat{n}$.  
The tensor formulation respects the full $RP^{2}$ symmetry of the 
uniaxial nematic since 
physical quantities, such as correlations, are written in terms $\Q$ which
is invariant under the local inversion $\hat{n} (\vec{r},t) 
\rightarrow - \hat{n} (\vec{r},t)$. At a string core 
$B = 3 A > 0$ and the eigensubspace corresponding to the largest  
eigenvalue $2 A/3$ is two-fold degenerate. Thus in the plane perpendicular to 
$\hat{h}$,
the tangent to the string, the orientation of the
director is ambiguous. At the isotropic core of a monopole $A=B=0$ and
all three 
eigenvalues of $\Q$ are degenerate so the orientation of the director is 
completely unspecified.

\begin{figure}
\makebox[\textwidth]{\epsfysize=4in \epsffile{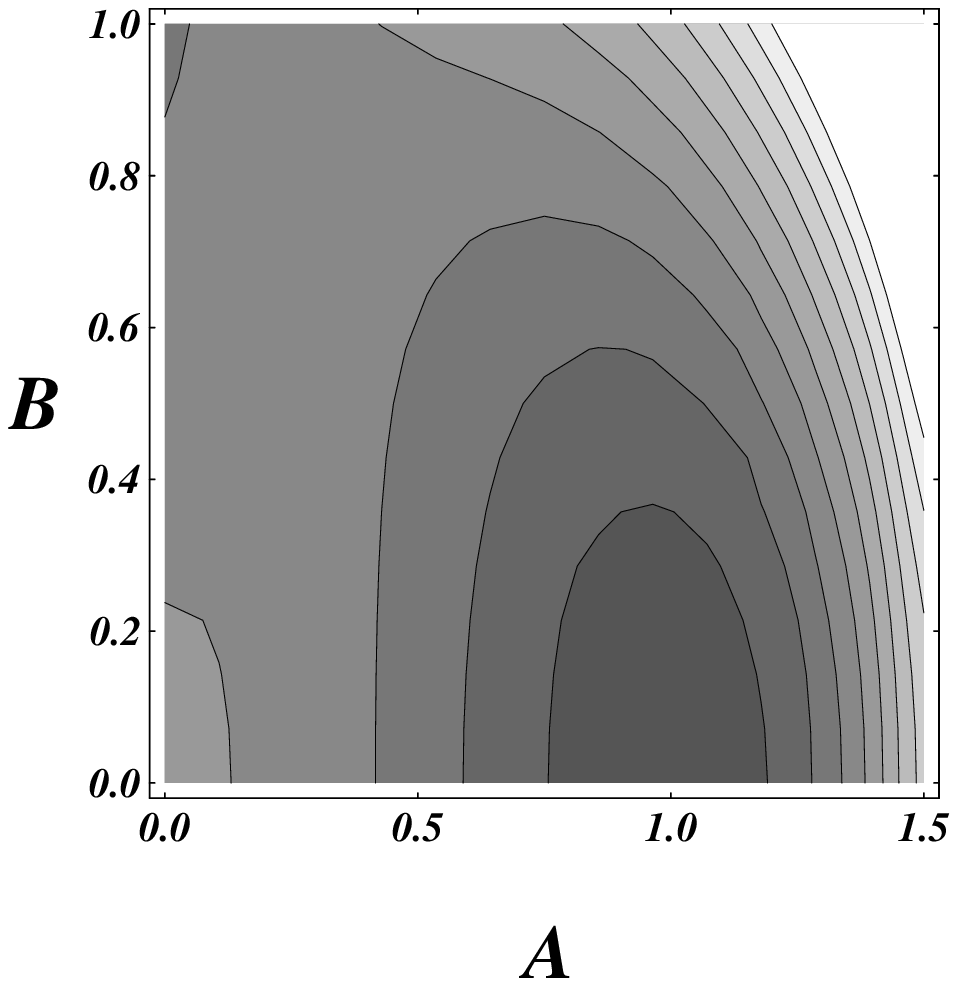}}
\caption{Contour plot of the potential surface $V(A,B)$ (\ref{EQ:VAP}) 
         for the bulk nematic 
         in the physical region $A>0$, $B>0$. The uniaxial amplitude is $A$
         and $B$ is the biaxial amplitude. Darker shades indicate lower 
         regions on the potential surface. The isotropic maximum at 
         $(A,B) = (0,0)$ corresponds to a monopole core, the saddle at 
         $(A,B) = (1/4,3/4)$ to a string core, and the minimum at 
         $(A,B) = (1,0)$ to the bulk uniaxial state. }
\label{FIG:POTENTIAL}
\end{figure}

The dynamics of the nematic is governed by the TDGL equation for $\Q$
%
% TDGL FOR NEMATICS
%
\be
\label{EQ:NEMTDGL}
\partial_{t} \Q = - \frac{\delta F[Q]}{\delta \Q} + \lambda_{\alpha
\beta} TrQ
\ee
with the Lagrange multiplier $\lambda_{\alpha \beta}$ included to enforce the
traceless condition. The Landau-de Gennes free energy is
%
% FREE - ENERGY
%
\be
\label{EQ:NEMFREEENERGY}
F[Q] = \int d^{3} r \mbox{ } [ \frac{1}{2} (\nabla Q)^{2} + V(Q) ]
\ee
with the potential
%
% POTENTIAL ENERGY
%
\be
\label{EQ:NEMPOTENTIAL}
V(Q) = - \frac{1}{6} Tr Q^{2} - \frac{1}{3} Tr Q^{3} + 
         \frac{1}{4} (Tr Q^{2})^{2}. 
\ee
The coefficient of the quadratic term in (\ref{EQ:NEMPOTENTIAL}) is chosen to
be negative so that the bulk isotropic phase is unstable towards nematic
ordering. The gradient term in (\ref{EQ:NEMFREEENERGY}) is written within the
equal-constant approximation \cite{DEGENNES}. Substitution of
the form (\ref{EQ:3DQ}) in (\ref{EQ:NEMPOTENTIAL}) results in a useful 
expression for the potential as a function of $A$ and $B$:
%
% FORM OF V(A,P)
%
\be
\label{EQ:VAP}
V(A,B) = - \frac{1}{9} A^{2} - \frac{2}{27} A^{3} + \frac{1}{9} A^{4} 
         - \frac{1}{27} B^{2} + \frac{1}{81} B^{4} + \frac{2}{27} [ A B^{2} 
         + A^{2} B^{2}].
\ee
A contour plot of $V(A,B)$ for $A>0$, $B>0$ is shown in 
Fig. \ref{FIG:POTENTIAL}. There 
is a global isotropic maximum at $(A,B) = (0,0)$, 
a uniaxial minimum at $(A,B) = (1,0)$, and a saddle at $(A,B) =
(1/4,3/4)$. The minimum represents the bulk nematic phase, 
the isotropic maximum corresponds to the monopole core, 
and the saddle, with $B = 3A$, signifies the string core.

Substituting (\ref{EQ:NEMFREEENERGY}) into
(\ref{EQ:NEMTDGL}) and using $TrQ = 0$ to calculate the Lagrange 
multiplier gives an explicit form for the equation of motion
%
% EXPLICIT EQUATION OF MOTION
%
\be
\label{EQ:NEMEOM}
\partial_{t} \Q = \nabla^{2} \Q - P_{\alpha \beta}
\ee
with the non-linear piece given by
%
% NON-LINEAR PIECE
%
\be
\label{EQ:NLP}
P_{\alpha \beta} =  (Tr Q^{2} - \frac{1}{3}) \Q - 
[Q^2]_{\alpha \beta} + \frac{\delta_{\alpha \beta}}{3} Tr Q^{2}.
\ee
Static solutions to (\ref{EQ:NEMEOM}) satisfy the 
Euler-Lagrange equation 
%
% EULER LAGRANGE EQUATION WITH SPATIAL DERIVATIVES
%
\be
\label{EQ:NEMEULER1}
\nabla^{2} \Q =  P_{\alpha \beta}.
\ee
The order-parameter correlation function is defined as
%
% OP CORRELATION FUNCTION
%
\be
\label{EQ:NEMCOR}
C(\vec{r},t) = 
{\cal N} \langle Tr [\delta Q(\vec{r},t) \delta Q(0,t)] \rangle
\ee
where ${\cal N} = (\langle Tr [\delta Q(0,t) \delta Q(0,t)] 
\rangle)^{-1}$ is
a normalization factor and $\delta \Q = \Q - \langle \Q \rangle$. Both 
${\cal N}$ and $\langle \Q \rangle$ are constant at leading order in $1/t$.  
Thus, at late-times, equation (\ref{EQ:NEMEOM}) can be written as 
an equation for the evolution of order-parameter correlations
%
% EQUATION FOR OP CORRELATIONS
%
\be
\label{EQ:NEMCOR2}
\frac{1}{2} \partial_{t} C(\vec{r},t) = 
\nabla^{2} C(\vec{r},t) - {\cal N} \langle Tr [P(\vec{r},t) \delta 
Q(0,t)] \rangle.
\ee
Later, through a development that closely parallels that previously given for 
the $O(n)$ model, it will be shown how (\ref{EQ:NEMEULER1}) 
and (\ref{EQ:NEMCOR2}) lead to a scaling equation for order-parameter 
correlations in the nematic.
%
% STRINGS ONLY
% 
\section{String defects in the nematic}

At late-times the dominant defects in the bulk nematic are strings with 
topological charge 1/2. Many of the main features of phase-ordering  in the
bulk nematic are described by the model containing strings which is presented
in Sec. III.B below. 
%
% CHARGE 1/2 VORTICES
%
\subsection{Vortices in thin films}

To begin, a model applicable to nematic thin films where
the director is constrained to lie in a plane without breaking the
$\hat{n} \rightarrow -\hat{n}$ symmetry is examined. 
By restricting the director to 
a plane, the intricacies of how to map the order-parameter tensor onto an 
auxiliary field when the director rotates by only $\pi$ about the vortex 
can be demonstrated, without the additional complication of biaxiality 
that appears near the string core in bulk samples.

For a uniaxial thin film nematic the order-parameter is a 
$2 \times 2$ traceless symmetric tensor 
%
% TWO-COMPONENT ORDER PARAMETER
%
\be
\label{EQ:FILMOP}
\Q = A [ \N_{\alpha} \N_{\beta} - \frac{1}{2} \delta_{\alpha \beta} ]
\ee
where $\hat{n}$ is the two-component director.
In analogy to the theory of the $O(2)$ model, the defects are incorporated
through a mapping of 
the order-parameter tensor onto a two-component auxiliary 
field. The only defect species present at late-times are charge $1/2$ 
point vortices with property that the director rotates by only $\pi$ around
the vortex. This property is essential in constructing the mapping. 
Consider a charge 
$1/2$ vortex at the origin with the typical director configuration
%
% PHYSICAL DIRECTOR FIELD
%
\be
\label{EQ:PHYSDIRECTOR}
\N = \cos \frac{1}{2} \phi \mbox{ } \hat{x} + \sin \frac{1}{2} \phi 
\mbox{ } \hat{y}
\ee
where $\phi$ is the polar angle in the $x-y$ plane. For 
future convenience we write the radial vector in the $x-y$ plane as 
$\SV$ and define angles in terms of $\hat{s}$ through
%
% S DEFN
%
\be
\label{EQ:S}
\hat{s} \equiv (\hat{s}_{1},\hat{s}_{2}) \equiv (\cos \phi,\sin \phi).
\ee
With the definitions (\ref{EQ:PHYSDIRECTOR}) and (\ref{EQ:S})  
the order-parameter tensor (\ref{EQ:FILMOP}) is \cite{DEEM96} 
%
% TYPE 1/2 OP TENSOR
%
\be
\label{EQ:OPS}
Q =  \frac{A(s)}{2} \left[ \begin{array}{cc}
\hat{s}_{1} & \hat{s}_{2} \\ \hat{s}_{2} & - \hat{s}_{1} \end{array} \right]
\ee
where $s =|\vec{s}|$. This form for $\Q$, analogous to the mapping 
(\ref{EQ:SIGEX}) for the $O(n)$ model, is a solution to the Euler-Lagrange 
equation (\ref{EQ:NEMEULER1}) written in terms of $s$
%
% FILM EULER LAGRANGE EQUATION
%
\be
\label{EQ:ELFILM}
\nabla^{2}_{s} \Q = \tilde{P}_{\alpha \beta}
\ee
where $\tilde{P}_{\alpha \beta}$ has a slightly modified definition from
$P_{\alpha \beta}$ (\ref{EQ:NLP}) because $\Q$ is a $2 \times 2$ tensor:
%
% MODIFIED P
%
\be
\tilde{P}_{\alpha \beta} =  (Tr Q^{2} - 1/3) \Q - 
[Q^2]_{\alpha \beta} + \frac{\delta_{\alpha \beta}}{2} Tr Q^{2}.
\ee
Substituting (\ref{EQ:OPS}) into (\ref{EQ:ELFILM}) results in an equation 
for the amplitude $A$:
%
% AMPLITUDE1/2
%
\be
\label{EQ:AMPLITUDE1/2}
\nabla_{s}^{2} A - \frac{A}{s^{2}} = 2 \frac{\partial U}{\partial A} (A).
\ee
From (\ref{EQ:NEMPOTENTIAL}) and (\ref{EQ:OPS}) the potential 
$U$ is given by
%
% POTENTIAL
%
\be
U(A) = - \frac{1}{12} A^{2} + \frac{1}{16} A^{4}.
\ee
An examination of (\ref{EQ:AMPLITUDE1/2}) at small $s$ has $A \sim s$, 
indicative of charge $1/2$ vortices \cite{CHARGE}.
At large $s$ the amplitude $A$ algebraically approaches its ordered value 
$A=\sqrt{2/3}$.

To treat many such vortices in a phase-ordering context,
$\vec{s}$ in (\ref{EQ:OPS}) is taken to be a Gaussian  field 
$\vec{s}(\vec{r},t)$ with zero mean. 
As in the $O(2)$ model, $s$ represents the distance 
to the nearest vortex, growing as the characteristic vortex spacing $L_{v}(t)$ 
at late-times. However, unlike in the $O(2)$ model, the director is not mapped 
directly onto $\SV$ - a $2 \pi$ rotation of $\SV$ about a vortex corresponds 
to a rotation of the director by $\pi$.

At late-times, the amplitude $A$ approaches it's ordered value and from the 
definition (\ref{EQ:NEMCOR}) and equation (\ref{EQ:OPS})
order-parameter correlation function is seen to be
%
% OP CORRELATION FOR 1/2 VORTICES
%
\be
C(\vec{r},t)  =  \langle \hat{s} (\vec{r},t) \cdot \hat{s} (0,t) \rangle
\ee
to leading order in $L_{v}^{-1}$.
This is just the $O(2)$ correlation function (\ref{EQ:OOOP}), and is
related to $f$, the correlation function for the auxiliary field $\SV$
defined in analogy to (\ref{EQ:DEFNF}), through (\ref{EQ:GAUSSIANC}) 
and (\ref{EQ:ONSCALF}) for $n=2$. In the 
scaling regime the equation of motion (\ref{EQ:NEMCOR2}) for $C(\vec{r},t)$
becomes equation (\ref{EQ:G-EQNF}) for the $O(2)$ scaling function $\F$, 
expressed in terms of the scaled length $x = r/L_{v}(t)$. The length
$L_{v}$ has the same definition as the length $L$ in (\ref{EQ:DEFNMU}),
with $\M$ replaced by $\SV$.
The path from  (\ref{EQ:NEMCOR2}) to (\ref{EQ:G-EQNF}) is similar to 
that taken in the $O(2)$ case \cite{LIU92a}.
The Euler-Lagrange equation (\ref{EQ:ELFILM}) is used to replace 
$\tilde{P}_{\alpha \beta}$ occuring in the last term in (\ref{EQ:NEMCOR2}).
The resulting expression is evaluated using the Gaussian identity
%
% GAUSSIAN IDENTITY
% 
\be
\label{EQ:GIFILM}
\langle Tr[ \nabla_{s}^{2} Q (\SV(\vec{r},t)) Q (\SV(0,t))] \rangle =
- \frac{2 f(\vec{r},t)}{\langle  [\SV(0,t)]^{2} \rangle} 
\frac{\partial}{\partial f(\vec{r},t)} \langle Tr Q [(\SV (\vec{r},t)) 
Q (\SV (0,t))] \rangle,
\ee
analogous to (\ref{EQ:GI}), and produces the last term on the
left-hand side of  (\ref{EQ:G-EQNF}).

Thus the scaling function $\F$ for the order-parameter correlations and the
growth law $L_{v} (t) \sim t^{1/2}$ for the nematic thin film 
are exactly those of the two-dimensional $O(2)$ model. This correspondence,
seen in simulations, can be simply understood as a consequence of the 
mapping of variables $\phi \rightarrow 2 \phi$ between the two models 
\cite{BLUNDELL92,DESAI}. This isomorphism
 is relevant to experimental efforts that
use constrained nematics to study coarsening in the two-dimensional 
$O(2)$ models \cite{XY} since it indicates that the existence of
the local $\hat{n} \rightarrow -\hat{n} $ symmetry does not affect
the leading order dynamics in the scaling regime.
%
% STRINGS IN THE BULK NEMATIC
%
\subsection{Strings in the bulk nematic}

In addition to the complication of a having director 
configuration with a charge $1/2$ 
geometry, strings in a bulk nematic have a biaxial core.
The form (\ref{EQ:3DQ}) for $\Q$ contains the biaxiality that is required if
an analytical solution to (\ref{EQ:NEMEULER1}) is to be found. String
defects enter the theory through the mapping of (\ref{EQ:3DQ}) 
onto a two-component auxiliary field. 
To motivate the form for the mapping consider the geometry of the 
director field around a charge $1/2$ string defect oriented along the 
$z$ axis. 
Since locally the coordinate system can always be chosen so that the string 
has this geometry the following development is quite general.
The director $\N$ is still given by (\ref{EQ:PHYSDIRECTOR}). The other
members of the orthonormal triad in (\ref{EQ:3DQ}) are 
%
% BIAXIAL PIECES
%
\bea
\hat{g} & = & -\sin \frac{1}{2} \phi \mbox{ } \hat{x} 
+ \cos \frac{1}{2} \phi \mbox{ } \hat{y} \\
\hat{h} & = & \hat{z}. 
\eea
With the notation (\ref{EQ:S}) for the radial vector $\vec{s}$ in the $x-y$
plane, the order-parameter tensor (\ref{EQ:3DQ}) becomes 
%
% COORDINATE REPRESENTATION OF THE ORDER PARAMETER
%
\be
\label{EQ:3DSTRING}
Q = \frac{A(s)}{2}  \left[ 
\begin{array}{ccc} 
\frac{1}{3} + \hat{s}_{1} & \hat{s}_{2} & 0 \\ 
\hat{s}_{2} & \frac{1}{3} - \hat{s}_{1} & 0 \\ 0 & 0 & - \frac{2}{3} 
\end{array}
 \right] 
+ \frac{B(s)}{6}  \left[ \begin{array}{ccc} 1 - \hat{s}_{1}  & - \hat{s}_{2} 
& 0 \\ - \hat{s}_{2} & 1 + \hat{s}_{1} & 0 \\ 0 & 0 & -2 
\end{array} \right].
\ee
This form for $\Q$ is a solution of (\ref{EQ:NEMEULER1}) written in terms 
of $s$,
%
% EULER IN TERMS OF S
%
\be
\label{EQ:EULERSTR}
\nabla_{s}^{2} \Q = P_{\alpha \beta}, 
\ee
provided that
%
% EQUATIONS FOR A AND P
%
\bea
4 \nabla_{s}^{2} A - \frac{(3 A - B)}{s^{2}} 
- 6  \frac{\partial V}{\partial A} & = & 0  \label{EQ:AB1} \\
\frac{4}{3} \nabla_{s}^{2} B + \frac{1}{3} \frac{(3 A - B)}{s^{2}}
- 6 \frac{\partial V}{\partial B} & = & 0, \label{EQ:AB2}
\eea
where $V(A,B)$ is given in (\ref{EQ:VAP}).
Note that equations (\ref{EQ:AB1}) and (\ref{EQ:AB2}) would be 
inconsistent had a 
uniaxial ansatz ($B=0$) been assumed at the outset. For the potential 
(\ref{EQ:VAP}) these equations are degenerate \cite{POTENTIAL}
and reduce to a single equation for $A$ after the identification $B = 1 - A$:
%
% EQUATION FOR A
% 
\be
\label{EQ:STRA}
4 \nabla_{s}^{2} A - \frac{(4 A - 1)}{s^{2}} 
- 6  \frac{\partial V}{\partial A} (A,1-A) =  0.
\ee 
At small $s$ the solution to (\ref{EQ:STRA}) is 
%
% SHORT DISTANCE STRING
%
\bea 
A & = & \frac{1}{4} + c s  \label{EQ:ASHORT} \\
B & = & \frac{3}{4} - c s  \label{EQ:PSHORT}
\eea
where $c$ is a constant, determined numerically.
At large $s$ the solution of (\ref{EQ:STRA}) takes the form
%
% LONG DISTANCE STRING
%
\bea
A & = & 1 - \frac{3}{4 s^{2}} \label{EQ:ALONG} \\
B & = &  \frac{3}{4 s^{2}}.   \label{EQ:PLONG}
\eea
As expected, the mapping (\ref{EQ:3DSTRING}) connects the biaxial 
saddle point on the potential surface $V(A,B)$, representing the string core,
to the uniaxial nematic minimum away from the string (see Fig. 
\ref{FIG:POTENTIAL}). The linear behaviour in (\ref{EQ:ASHORT}) and 
(\ref{EQ:PSHORT}) at small $s$ is that expected for charge $1/2$ strings
in the nematic. Both the linear behaviour near the core and the algebraic 
relaxation (\ref{EQ:ALONG}-\ref{EQ:PLONG}) to the bulk uniaxial state
are seen in the numerical results of \cite{SCHOPOHL87}. 

Once again, to examine the statistical properties of the string defect tangle,
$\vec{s}$ is a taken to be a Gaussian auxiliary field with zero mean.
The magnitude $s$ grows
as the characteristic string separation $L_{s}(t)$. Therefore, at late-times,
$s$ is large and the biaxial piece of $\Q$, with an amplitude $B$ given by 
(\ref{EQ:PLONG}), is suppressed. This is physically reasonable since 
biaxiality occurs on length scales around the core size 
while the late-time scaling 
properties are dominated by physics at the much larger scale of $L_{s}(t)$.
At late-times, when $A \approx 1$, the definition (\ref{EQ:NEMCOR}) and the
mapping (\ref{EQ:3DSTRING}) show that the order-parameter correlation 
function reduces to 
%
% STRING CORRELATION FUNCTION
%
\be
C(\vec{r},t) = \langle \hat{s} (\vec{r},t) \cdot \hat{s} (0,t) \rangle
\ee
which is the $O(2)$ correlation function (\ref{EQ:OOOP}). As before, 
$C(\vec{r},t)$ is related to $f(\vec{r},t)$, the normalized correlation 
function for the auxiliary field $\SV$, 
by relations (\ref{EQ:GAUSSIANC}) and (\ref{EQ:ONSCALF}) with $n=2$. 

The dynamical
equation (\ref{EQ:NEMCOR2}) for $C(\vec{r},t)$ reduces, in the scaling regime,
to (\ref{EQ:G-EQNF}) for $\F$ from the three-dimensional $O(2)$ model.
Note that the spatial dimensionality enters through the Laplacian operator in 
(\ref{EQ:G-EQNF}).
The scaled length in this case is $x=r/L_{s}(t)$ with $L_{s}$ defined 
as $L$ in (\ref{EQ:DEFNMU}).  The derivation of this correspondence parallels
the steps taken in the $O(2)$ model that lead to (\ref{EQ:G-EQNF}).
The Euler-Lagrange equation
(\ref{EQ:EULERSTR}) enables the non-linear quantity 
$P_{\alpha \beta}$ occuring in the last term of (\ref{EQ:NEMCOR2}) to 
be replaced by $\nabla_{s}^{2} \Q$. The resulting average 
is then evaluated using (\ref{EQ:GIFILM})
and produces the last term on the left-hand side 
of
(\ref{EQ:G-EQNF}).
The single-length scaling result $L_{s} \sim t^{1/2}$ is recovered for 
the
phase ordering of the bulk nematic. In Fig. \ref{FIG:STR} the 
theoretical 
results for $\F$ in the the three-dimensional $O(2)$ model 
\cite{LIU92a} 
and the $\F$ determined in cell-dynamical simulations of the 
bulk nematic 
\cite{BLUNDELL92} are compared. The agreement between the two is 
excellent. 
At short-scaled  distances $\F$ has the form (\ref{EQ:GS}) which 
is also 
seen in the simulations and is an indication that string defects 
are the 
dominant disordering agent in the bulk nematic.

The theory is now structured so that many well-established 
phase-ordering results for the $O(2)$ model \cite{LIU92a,LIU92b} can be
applied to the bulk nematic.
In particular, the string line density $\eta$ is related to the 
auxiliary field $\SV$, whose
zeros locate the positions of the strings,  through 
\cite{LIU92b,WICKHAM97b,TOYOKI87}
%
% STRING DENSITY 
%
\be
\label{EQ:STRDEN}
\eta = \delta(\SV) |\vec{\omega}|
\ee
where the tangent to the string,
%
% OMEGA
%
\be
\label{EQ:STROMEGA}
\vec{\omega} = \nabla s_{1} \times \nabla s_{2}, 
\ee
points in the direction of positive winding number. The calculation 
performed in Appendix A
shows that the average line density of string obeys 
$\langle \eta \rangle \sim L_{s}^{-2} \sim t^{-1}$ for late-times,
in accord with experiments \cite{CHUANG93}.
%
% MONOPOLES ONLY
%
\section{Monopoles in the bulk nematic}

To address experiments that are designed to produce copious amounts of 
monopoles at the quench \cite{PARGELLIS96}, a theory for
the ordering kinetics of bulk nematics is considered in which monopoles appear.
The model consists of mapping the director $\N$ near a monopole
directly onto a three-component Gaussian auxiliary field $\M$
{\em via} $\N = \hat{m}$. Thus the order-parameter is 
%
% NAIVE ORDER PARAMETER
%
\be
\label{EQ:NAIVEOP}
\Q = A(m) [ \HM_{\alpha} \HM_{\beta} - \frac{1}{3} \delta_{\alpha \beta} ].
\ee
Since the isotropic monopole core can be connected to the nematic minimum 
along the $B=0$ line on the potential surface (Fig. \ref{FIG:POTENTIAL}), a
biaxial piece does not appear in (\ref{EQ:NAIVEOP}). Equation 
(\ref{EQ:NAIVEOP}) solves the Euler-Lagrange equation 
(\ref{EQ:NEMEULER1}) written in terms of $\M$,
%
% EULER-LAGRANGE WITH M
%
\be
\label{EQ:MONEL}
\nabla^{2}_{m} \Q = P_{\alpha \beta},
\ee
if the amplitude $A$ satisfies
%
% AMPLITUDE EQUATION
%
\be
\label{EQ:MONAMP}
\nabla^{2}_{m} A - \frac{6}{m^2} A = \frac{3}{2} 
\frac{\partial V}{\partial A} (A,0).
\ee
A similar result was obtained in \cite{SCHOPOHL88} for equilibrium. 
For small $m$, (\ref{EQ:MONAMP}) indicates that $A \sim m^{2}$ while for 
large $m$ the amplitude $A$ algebraically approaches its ordered value of $1$.
The $m^{2}$ dependence at small $m$ indicates that (\ref{EQ:NAIVEOP}) 
describes charge 1 monopoles in the nematic \cite{CHARGE}. This is also 
evident geometrically since $\M$ (and thus $\hat{n}$) 
is a radial vector field near the monopole.

At late-times, using (\ref{EQ:NAIVEOP}) with $A \approx 1$ 
the order-parameter correlation function (\ref{EQ:NEMCOR}) is  
%
% OP1
%
\be
\label{EQ:OP1}
C(\vec{r},t) = \frac{3}{2} \left[ \langle [\HM(\vec{r},t) \cdot \HM(0,t) 
]^{2} \rangle -\frac{1}{3} \right].
\ee
In contrast to the string models considered earlier, the 
expression (\ref{EQ:OP1}) for the order-parameter correlation function in 
the monopole model is new.
The Gaussian average in (\ref{EQ:OP1}) is computed in Appendix B. In the
late-time scaling regime $C(\vec{r},t)$ can be 
written in terms of the scaled length $x = r/L_{m}(t)$, where $L_{m}(t)$ 
is the typical monopole separation. Thus $C(\vec{r},t) = \F (x)$ with 
%
% FORM FOR C
%
\be
\label{EQ:NAIVESCALF}
\F = 1 + \frac{3}{\gamma^{3} f^{3}} ( \sin^{-1} f - \gamma f)
\ee
and $\gamma = 1/\sqrt{1-f^{2}}$. The auxiliary field correlation function $f$
is defined in (\ref{EQ:DEFNF}).
The scaling function $\F$ 
satisfies the scaling equation (\ref{EQ:G-EQNF}) with $L_{m}(t) \sim t^{1/2}$.
The development of this result closely parallels that of the string case 
considered earlier.
The only difference between the scaling results for this model and those
for the $O(3)$ model is that the relation between 
$\F$ and $f$ is (\ref{EQ:NAIVESCALF}) instead of (\ref{EQ:ONSCALF}).

Since $\M$ is smooth, $f$ has power series expansion that is analytic at 
small $x$. By using this expansion in (\ref{EQ:G-EQNF}) and 
(\ref{EQ:NAIVESCALF}) the small-$x$ behaviour of $\F$ is found to be
%
% SHORT DISTANCE FORM FOR NNM
%
\be
\label{EQ:NEMSHORT}
\F(x) =  1 - \frac{3 \pi}{2 \mu d} x^{2} + 
            \frac{3 \pi^{2} }{4 \mu (d+1)}\sqrt{ \frac{\pi}{2 \mu d}}
\mbox{ } x^{3} + {\cal O} (x^{4}).
\ee
The non-analytic $x^{3}$ term in $\F$, also found in the $O(3)$ model 
(\ref{EQ:GS1}),  
is due to the presence of point monopole defects. Using a fourth-order 
Runge-Kutta scheme, 
the non-linear eigenvalue problem represented by
(\ref{EQ:G-EQNF}) and (\ref{EQ:NAIVESCALF}) is solved 
for $d=3$. The eigenvalue is 
$\mu = 1.27306 \dots$, which differs from the value $\mu = 0.5558 \dots$ 
for the 
$O(3)$ model \cite{LIU92a}. The function $\F$ is plotted in Fig. 
\ref{FIG:MON} along with the scaling function for order-parameter correlations
in the three-dimensional
$O(3)$ model. Fig. \ref{FIG:MON} also compares the 
cell-dynamical simulation data for the bulk nematic \cite{BLUNDELL92} 
to the function $\F$, equation (\ref{EQ:NAIVESCALF}). 
The function $\F$ does not describe the simulation 
data as well as the string model, showing deviations at short distances.
These deviations are expected since the structure of the theory
at short distances
(\ref{EQ:NEMSHORT}) represents
the wrong defects (monopoles) instead of the correct ones (strings).

Since the zeros of $\M$ locate the monopole cores,
the monopole number density $n$ can be expressed in terms of
the auxiliary field $\M$ \cite{LIU92b} as 
%
% MONOPOLE DENSITY
% 
\be
\label{EQ:MONDEN}
n = \delta(\M) \mbox{ $|$det}( \partial_{i} m_{j})|.
\ee
where the quantity between the absolute value signs is the Jacobian for the
transformation from real space coordinates to auxiliary field variables.
From the development in \cite{LIU92b} the average monopole number density obeys
$\langle n \rangle \sim L_{m}^{-3} \sim t^{-3/2}$. This result holds only
for monopole annihilation in the absence of strings, the case considered in
this section. In the experiments of Chuang {\em et al.} \cite{CHUANG93}
where monopole annihilation occured in 
the presence of strings the average monopole density was observed to decay 
faster, with 	
$\langle n \rangle \sim t^{-3}$.
%
% DISCUSSION
%
\section{discussion}

The dominant scaling behaviour observed during ordering in the bulk nematic is 
well-described by the the theory presented here in which
string defects are the major disordering agents. The growth law 
$L_{s} \sim t^{1/2}$ is recovered, leading to an average 
 string line density $\langle \eta \rangle$
that decays as $\langle \eta \rangle \sim t^{-1}$, as seen in experiments 
\cite{CHUANG93}. The theoretically determined scaling form for order-parameter 
correlations in the bulk nematic is shown analytically to be {\em exactly} 
that for the three-dimensional $O(2)$ model \cite{LIU92a}, 
and this is in excellent agreement with the simulation results 
\cite{BLUNDELL92} (Fig. \ref{FIG:STR}).  This paper addresses the 
issue of biaxiality near the string cores and demonstrates that
it is irrelevant to the leading order scaling properties of the system. 
However, the theory is capable of being extended into the pre-scaling regime, 
where biaxiality may play a role in the dynamics. 

The major accomplishment of this work is the explicit demonstration 
of the isomorphism between the late-stage ordering in the bulk nematic 
and the late-stage ordering in the three-dimensional $O(2)$ model, within the
Gaussian approximation. It is 
shown that, in the scaling regime, the order-parameter equations of 
motion for the $O(2)$ model (\ref{EQ:TDGL}) and the bulk nematic 
(\ref{EQ:NEMTDGL}) produce the same scaling equation (\ref{EQ:G-EQNF})
for the correlation function.  The essential element in the present theory, 
which was missing in earlier theories \cite{BRAY93}, is the mapping 
(\ref{EQ:3DSTRING}), which explicitly includes string defects
and makes a direct connection with the $O(2)$ model. 
As a consequence, results for the $O(2)$ model, such as
string and vortex density correlations \cite{LIU92b,MAZENKO97VA} or
conservation laws involving string densities \cite{MAZENKO97OJK}, can be
directly applied to the bulk nematic.

This paper also presents a model for bulk nematics in which monopoles appear.
The model is applicable to situations where monopole-antimonopole 
annihilations occur in isolation from string defects. Such scenarios have been
realized experimentally \cite{PARGELLIS96}, and the data is suggestive of the
growth law $L_{m} \sim t^{1/2}$ predicted by the theory. However, to properly 
treat monopole dynamics in the presence of strings, theories that include
interactions between string and monopole defects are required. This 
interesting aspect of the problem is under current investigation 
\cite{WICKHAM97b}.
%
% ACKNOWLEDGEMENTS
%
\acknowledgements

The author thanks Gene Mazenko for guidance and for many stimulating 
discussions. The author also benefited from discussions with 
Alan Bray, Andrew Rutenberg, Bernard Yurke, and Martin Zapotocky. The 
simulation data shown in this paper was graciously provided by Rob Blundell.
Support from the NSERC of Canada is gratefully acknowledged. This work 
was supported in part by the MRSEC Program 
of the National Science Foundation under Award Number DMR-9400379.
%
% APPENDIX
%
\appendix
\section{}
This appendix presents the calculation of the average line density for 
strings, $\langle \eta \rangle$, for the $O(n)$ model in $d=n+1$ spatial 
dimensions, defined as
\be
\label{EQ:ASLD}
\langle \eta \rangle = \langle \delta(\SV) |\vec{\omega}| \rangle
\ee
with 
\be
\label{EQ:ATAN}
\omega_{\alpha} = \frac{1}{n!} \epsilon_{\alpha \mu_{1} \dots \mu_{n}}
\epsilon_{\nu_{1} \dots \nu_{n}} \nabla_{\mu_{1}} s_{\nu_{1}} \dots 
\nabla_{\mu_{n}} s_{\nu_{n}}.
\ee
The form (\ref{EQ:ATAN}), where $\epsilon$ is the fully antisymmetric tensor,
generalizes the definition (\ref{EQ:STROMEGA}) for the tangent vector to a 
string in $d=3$. The one-point average (\ref{EQ:ASLD}) can be written in 
an integral form
%
% INTEGRAL FORM
%
\be
\label{EQ:AINT}
\langle \eta \rangle = \int \prod_{\mu=1,\nu=1}^{n+1,n} d 
\xi_{\mu}^{\nu} \mbox{ } |\vec{\omega}(\xi)| G(\xi)
\ee
in terms of
%
% OMEGA
%
\be
\omega_{\alpha} (\xi) = \frac{1}{n!} \epsilon_{\alpha \mu_{1} \dots \mu_{n}}
\epsilon_{\nu_{1} \dots \nu_{n}} \xi_{\mu_{1}}^{\nu_{1}} \dots 
\xi_{\mu_{n}}^{\nu_{n}}
\ee
and the one-point reduced probability distribution 
$G(\xi)$, given by 
%
% G
%
\be
\label{EQ:ARPD}
G(\xi) = \langle \delta(\SV) \prod_{\mu=1,\nu=1}^{n+1,n} \delta(\xi_{\mu}^{\nu}
- \nabla_{\mu} s_{\nu} ) \rangle.
\ee
The Gaussian average in (\ref{EQ:ARPD})
is straightforward to evaluate by first writing the 
$\delta$-functions in the integral representation and then performing the 
resulting standard Gaussian integrals. One finds
%
% FORM FOR G
%
\be
\label{EQ:AGFIN}
G(\xi) = \frac{1}{(2 \pi S_{0}(t))^{n/2}} \frac{1}{(2 \pi S^{(2)})^{n(n+1)/2}}
\exp - \sum_{\mu=1,\nu=1}^{n+1,n} \frac{(\xi_{\mu}^{\nu})^{2}}{2 S^{(2)}}
\ee
with the definitions
\bea
S_{0}(t) & = & \frac{1}{n} \langle [\SV (0,t) ]^{2} \rangle \\
S^{(2)} & = & \frac{1}{n(n+1)} \sum_{\mu=1,\nu=1}^{n+1,n} 
\langle [\nabla_{\mu} s_{\nu}]^{2} \rangle.
\eea
In this theory $S^{(2)} = 1/(n+1)$ \cite{WICKHAM97b}. 
Substitution of (\ref{EQ:AGFIN}) in (\ref{EQ:AINT}) produces the final form
for the average line density of string:
%
% RESULT
%
\be
\label{EQ:FINAL}
\langle \eta \rangle = C_{n} \left[ \frac{S^{(2)}}{\pi S_{0}(t)} \right]^{n/2}
\ee
with the $n$-dependent constant $C_{n}$ defined as
\be
C_{n} = \frac{1}{\pi^{n(n+1)/2}} \int  \prod_{\mu=1,\nu=1}^{n+1,n} d 
\xi_{\mu}^{\nu} \mbox{ }|\vec{\omega}(\xi)| \exp - \sum_{\mu=1,\nu=1}^{n+1,n} 
(\xi_{\mu}^{\nu})^{2}.
\ee
For $n=2$ it can be shown that $C_{2} = 1$ \cite{WICKHAM97b}.
Since  $S_{0}(t) \sim t$ at late-times, 
the average line density of string scales like
\be
\langle \eta \rangle \sim t^{-n/2}.
\ee
In particular, for $n=2$, $\langle \eta \rangle \sim t^{-1}$.
\section{}
This appendix outlines the evaluation of the average
%
% AVERAGE
%
\be
A = \langle [\hat{m} (\vec{r},t) \cdot \hat{m} (0,t) ]^{2} \rangle
\ee
appearing in the correlation function (\ref{EQ:OP1}) for the monopole model.
For an $n$-component Gaussian $\M$ field, the average $A$ can be written in 
the integral form 
\be
A = \int d^{n} x_{1} d^{n} x_{2} \frac{(\vec{x}_{1} \cdot \vec{x}_{2})^{2}}
{(x_{1})^{2} (x_{2})^{2}} \Phi
(\vec{x}_{1},\vec{x}_{2})
\ee
in terms of the two-point reduced probability distribution \cite{LIU92a}
\be
\Phi(\vec{x}_{1},\vec{x}_{2}) = \left[
\frac{\gamma}{2 \pi} \right]^{n}
\exp -\frac{\gamma^{2}}{2} (\vec{x}_{1}^{2} + \vec{x}_{2}^{2} 
- 2 f \vec{x}_{1} \cdot \vec{x}_{2} )
\ee
where the auxiliary field correlation function $f$ is defined in 
(\ref{EQ:DEFNF}) and $\gamma = 1/\sqrt{1-f^{2}}$.
The identity 
\be
\frac{1}{(x_{1})^{2}} = 2 \int_{0}^{\infty}
 d r_{1} r_{1} \exp - x_{1}^{2} r_{1}^{2}
\ee
allows $A$ to be written as a Gaussian integral
\be
\label{EQ:AAGAUSS}
A = \lim_{\lambda \rightarrow  1} \int_{0}^{\infty} d r_{1} r_{1}
\int_{0}^{\infty} d r_{2} r_{2} \frac{\partial^{2}}{\partial \lambda^{2}}
\int d^{n} x_{1} d^{n} x_{2} \mbox{ } 
\tilde{\Phi}_{\lambda} (\vec{x}_{1},\vec{x}_{2},r_{1},r_{2})
\ee
with
\be
\tilde{\Phi}_{\lambda} (\vec{x}_{1},\vec{x}_{2},r_{1},r_{2}) = 
\frac{4}{f^{2} \gamma^{4}} 
\left[\frac{\gamma}{2 \pi} \right]^{n}
\exp -(r_{1}^{2} + \frac{\gamma^{2}}{2} ) x_{1}^{2} - (r_{2}^{2} + 
 \frac{\gamma^{2}}{2} ) x_{2}^{2} + \gamma^{2} f \lambda 
\mbox{ } \vec{x}_{1} \cdot
\vec{x}_{2}. 
\ee
The integrals over $\vec{x}_{1}$ and $\vec{x}_{2}$ in (\ref{EQ:AAGAUSS})
are readily done. After
differentiating with respect to $\lambda$ and setting $\lambda=1$, the 
integral over $r_{1}$ is performed. After a change of variables, 
$y=(r_{2})^{2}$, the following integrals remain:
%
% INTEGRAL FORMS
%
\be
A = 2^{-(n/2+1)} \int_{0}^{\infty} dy (y + \frac{\gamma^{2}}{2})^{-1} 
(y + \frac{1}{2})^{-n/2} + 2^{-(n/2 + 2)} n \gamma^{2} f^{2} 
\int_{0}^{\infty} dy (y + \frac{\gamma^{2}}{2})^{-1} 
(y + \frac{1}{2})^{-(n/2+1)}.
\ee
These integrals can be expressed in terms of hypergeometric functions $F$
\cite{GRASSSTAIN}, giving
%
% FINAL FORM
%
\bea
A & = & 1 - \frac{\ln \gamma}{\gamma^{2} -1} \mbox{ for } n=2 \\
A & = & 1 + \frac{(n-1)}{\gamma^{2} (n-2)} 
\left[ \frac{1}{\gamma^{2} f^{2}} (F[1,1;n/2;f^{2}]-1) -1 \right] \mbox{ for }
n > 2 .
\label{EQ:NG2}
\eea
In particular, for $n=3$, equation (\ref{EQ:NG2}) gives 
\be
A =  1 + \frac{2}{\gamma^{2}} 
\left[ \frac{1}{\gamma^{2} f^{3}} (\gamma \sin^{-1} f -f)  - 1\right],  
\ee
which leads to (\ref{EQ:NAIVESCALF}) for $\F$ in the nematic with monopoles.
%
% BIBLIOGRAPHY
%

%
% FIGURES
\pagebreak
\begin{figure}
% GNUPLOT: LaTeX picture
\setlength{\unitlength}{0.240900pt}
\ifx\plotpoint\undefined\newsavebox{\plotpoint}\fi
\sbox{\plotpoint}{\rule[-0.200pt]{0.400pt}{0.400pt}}%

\vspace{.5 in}
\caption{The scaling function $\F(x)$ as a function of the scaled length
$x=r/L_{m}(t)$ for order-parameter correlations in the theory for 
monopoles in bulk nematics, discussed in Section IV is represented by the
lower curve. 
The upper curve is the scaling function for order-parameter correlations 
in the three-dimensional $O(3)$ model, from 
\protect\cite{LIU92a}. The circles  are the cell-dynamical simulation data of
 Blundell and Bray \protect\cite{BLUNDELL92} for $\F$ in a bulk nematic.
 The abscissa of the simulation data is scaled so as to 
give the best fit to the theory for monopoles in bulk nematics.}
\label{FIG:MON}
\end{figure}
\pagebreak

\begin{thebibliography}{99}
\bibitem{CHARGE}      This can be seen by examining 
                      the explicit expression for the charge
                      on a defect, given in \cite{LIU92b}. For small distances
                      $r$ from the defect core a charge $q$ defect has $A(r) 
                      \sim r^{q}$ in the $O(n)$ model. For a nematic this would
                      represent a charge $q/2$ defect, using the standard 
                      definitions \cite{DEGENNES}.
\bibitem{POTENTIAL}  This special state of affairs is due to a judicial choice
                     of coefficients in the potential 
                     (\ref{EQ:NEMPOTENTIAL}). However, the scaling  
                     properties  of the late-time regime are independent of the
                     details of the potential. Thus this particular choice is 
                     simply convenient - any potential with the same generic 
                     properties is expected to give the same scaling results.
\end{thebibliography}
\end{document}